\documentstyle[12pt,epsfig]{article}

\begin{document}
\begin{titlepage}
\begin{center}

{\Large\bf Role of Positivity Constraints in Determining\\[2mm]
Polarized Parton Densities}

\end{center}
\vskip 2cm
\begin{center}
{\bf Elliot Leader}\\
{\it Imperial College London\\ Prince Consort Road, London SW7
2BW, England }
%E-mail: e.leader@imperial.ac.uk}
\vskip 0.5cm
{\bf Aleksander V. Sidorov}\\
{\it Bogoliubov Theoretical Laboratory\\
Joint Institute for Nuclear Research, 141980 Dubna, Russia }
%E-mail: sidorov@thsun1.jinr.ru}
\vskip 0.5cm
{\bf Dimiter B. Stamenov \\
{\it Institute for Nuclear Research and Nuclear Energy\\
Bulgarian Academy of Sciences\\
Blvd. Tsarigradsko Chaussee 72, Sofia 1784, Bulgaria }}
%E-mail:stamenov@inrne.bas.bg
\end{center}

\vskip 0.3cm
\begin{abstract}
\hskip -5mm

We have re-analyzed the world data on inclusive polarized DIS, in
both NLO and LO QCD, including the very precise JLab Hall A
neutron data, and have studied the role of positivity constraints,
and demonstrated their importance, in determining the strange and
gluon densities. We have shown that higher twist corrections are
essential in the analysis of the present data on the structure
function $g_1$. A consistent QCD analysis is achieved, and results
for the polarized parton densities are given in both the $\rm
\overline{MS}$ and JET schemes.

\end{abstract}

\end{titlepage}

\newpage
\setcounter{page}{1}

\section{Introduction}

Spurred on by the famous European Muon Collaboration (EMC)
experiment \cite{EMC} at CERN in 1987, there has been a huge
growth of interest in the partonic spin structure of the nucleon,
{\it i.e.}, how the nucleon spin is built up out from the
intrinsic spin and orbital angular momentum of its constituents,
quarks and gluons. Our present knowledge about the spin structure
of the nucleon comes from polarized inclusive and semi-inclusive
DIS experiments at SLAC, CERN, DESY and JLab, polarized
proton-proton collisions at RHIC and polarized photoproduction
experiments. The determination of the longitudinal polarized
parton densities in QCD is one of the important aspects of this
knowledge.

In this paper we present an updated version of our NLO QCD
polarized parton densities in both the $\rm \overline{MS}$ and
the JET (or so-called chirally invariant) \cite{JET}
factorization schemes, as well as the LO ones, determined from the
world data \cite{EMC,world,JLab} on {\it inclusive} polarized DIS.
Comparing to our previous analyses \cite{LSS2001, LSSHT}: i) The
recent very precise JLab/ Hall A data \cite{JLab} on
$g_1^n/F_1^n$ are incorporated into the analysis and ii) New
positivity constraints are imposed and their role in the
determination of the polarized parton densities is discussed. The
updated polarized parton densities (PPD) are compared to those
obtained by the other groups and the effect of different
positivity constraints that have been imposed is demonstrated.

\section{QCD fits to the data}

In QCD the spin structure function $g_1$ can be written in the
following form ($Q^2 >> \Lambda^2$):
\begin{equation}
g_1(x, Q^2) = g_1(x, Q^2)_{\rm LT} + g_1(x, Q^2)_{\rm HT}~,
\label{g1QCD}
\end{equation}
where "LT" denotes the leading twist ($\tau=2$) contribution to
$g_1$, while "HT" denotes the contribution to $g_1$ arising from
QCD operators of higher twist, namely $\tau \geq 3$. In
(\ref{g1QCD}) we have dropped the nucleon target label N. The HT
power corrections (up to ${\cal O}(1/Q^2)$ terms) can be divided
into two parts:
\begin{equation}
g_1(x, Q^2)_{\rm HT}= h^{\rm TMC}(x, Q^2)/Q^2 + h(x, Q^2)/Q^2~,
\label{HTQCD}
\end{equation}
where $h^{\rm TMC}(x, Q^2)$ are the calculable {\cite{TB}
kinematic target mass corrections and effectively belong to the
LT term. $h(x, Q^2)$ are the {\it dynamical} higher twist
($\tau=3$ and $\tau=4$) corrections to $g_1$, which are related to
multi-parton correlations in the nucleon. The latter are
non-perturbative effects and cannot be calculated without using
models. (Note that the twist-3 contribution to $g_1$ is equal to
zero if the Wandzura-Wilczek approximation \cite{WW} for the spin
structure function $g_2$ is used.) $g_1(x, Q^2)_{\rm LT}$ in
(\ref{g1QCD}) is the well known pQCD expression and in NLO has
the form
\begin{equation}
g_1(x,Q^2)_{\rm pQCD}={1\over 2}\sum _{q} ^{N_f}e_{q}^2 [(\Delta
q +\Delta\bar{q})\otimes (1 + {\alpha_s(Q^2)\over 2\pi}\delta
C_q) +{\alpha_s(Q^2)\over 2\pi}\Delta G\otimes {\delta C_G\over
N_f}], \label{g1partons}
\end{equation}
where $\Delta q(x,Q^2), \Delta\bar{q}(x,Q^2)$ and $\Delta
G(x,Q^2)$ are quark, anti-quark and gluon polarized densities in
the proton, which evolve in $Q^2$ according to the spin-dependent
NLO DGLAP equations. $\delta C(x)_{q,G}$ are the NLO
spin-dependent Wilson coefficient functions and the symbol
$\otimes$ denotes the usual convolution in Bjorken $x$ space.
$\rm N_f$ is the number of active flavors. Note that in the LO QCD
approximation the coefficients $\delta C(x)_{q,G}$ in
(\ref{g1partons}) vanish and the polarized parton densities
evolve according the LO DGLAP equations.

One of the features of polarized DIS is that a lot of the present
data are in the preasymptotic region ($Q^2 \sim 1-5~GeV^2,
4~GeV^2 < W^2 < 10~GeV^2$). While in the unpolarized case we can
cut the low $Q^2$ and $W^2$ data in order to minimize the less
well known higher twist effects, it is impossible to perform such
a procedure for the present data on the spin-dependent structure
functions without losing too much information. This is especially
the case for the HERMES, SLAC and Jefferson Lab experiments. So,
to confront correctly the QCD predictions with the experimental
data and to determine the {\it polarized} parton densities
special attention must be paid to the non-perturbative higher
twist (powers in $1/Q^2$) corrections to the nucleon structure
functions.

We have used two approaches to extract the polarized parton
densities from the world polarized DIS data. According to the
first \cite{LSS2001} the leading twist LO/NLO QCD expressions for
the structure functions $g_1$ and $F_1$ have been used in order
to confront the data on spin asymmetry $A_1 (\approx g_1/F_1)$
and $g_1/F_1$. We will refer to these as '$g_1/F_1$' fits. We have
shown \cite{LomConf,newHTA1} that in this case the extracted from
the world data 'effective' HT corrections $h^{g_1/F_1}(x)$ to the
ratio $g_1/F_1$
\begin{equation}
\left[{g_1(x,Q^2)\over
F_1(x,Q^2)}\right]_{exp}~\Leftrightarrow~{g_1(x,Q^2)_{\rm LT}\over
F_1(x,Q^2)_{\rm LT}} + {h^{g_1/F_1}(x)\over Q^2} \label{A1HT}
\end{equation}
are negligible and consistent with zero within the errors, {\it
i.e.} $h^{g_1/F_1}(x) \approx 0$, when for $(g_1)_{LT}$ and
$(F_1)_{LT}$ their NLO QCD approximations are used. (Note that in
QCD the unpolarized structure function $F_1$ takes the same form
as $g_1$ in (\ref{g1QCD}), namely $F_1 = (F_1)_{LT} +
(F_1)_{HT}.$) What follows from this result is that the higher
twist corrections to $g_1$ and $F_1$ approximately compensate each
other in the ratio $g_1/F_1$ and the NLO PPDs extracted this way
are less sensitive to higher twist effects. This is not true in
the LO case (see our discussion in Ref. \cite{LSSHT}). The set of
polarized parton densities extracted this way is referred to as
PD($g_1^{\rm NLO}/F_1^{\rm NLO}$) or PD(Set 1).

According to the second approach \cite{LSSHT}, the $g_1/F_1$ and
$A_1$ data have been fitted using phenomenological
parametrizations of the experimental data for the unpolarized
structure function $F_2(x,Q^2)$ and the ratio $R(x,Q^2)$ of the
longitudinal to transverse $\gamma N$ cross-sections (i.e. $F_1$
is replaced by its expression in terms of usually extracted from
unpolarized DIS experiments $F_2$ and $R$). Note that such a
procedure is equivalent to a fit to $(g_1)_{exp}$, but it is more
consistent than the fit to the $g_1$ data themselves actually
presented by the experimental groups because here the $g_1$ data are
extracted in the same way for all of the data sets. In this case
the HT corrections to $g_1$ cannot be compensated because the HT
corrections to $F_1(F_2$ and $R)$ are absorbed in the
phenomenological parametrizations of the data on $F_2$ and $R$.
Therefore, to extract correctly the polarized parton densities
from the $g_1$ data, the HT corrections (\ref{HTQCD}) to $g_1$
have to be taken into account. In our fit to the data we have
used the following expressions for $g_1/F_1$ and $A_1$:
\begin{eqnarray}
\nonumber \left[{g_1^N(x,Q^2)\over F_1^N(x,
Q^2)}\right]_{exp}~&\Leftrightarrow&~ {{g_1^N(x,Q^2)_{\rm
LT}+h^N(x)/Q^2}\over F_2^N(x,Q^2)_{exp}}2x{[1+
R(x,Q^2)_{exp}]\over (1+\gamma^2)}~,\\
A_1^N(x,Q^2)_{exp}~&\Leftrightarrow&~{{g_1^N(x,Q^2)_{\rm LT}+
h^N(x)/Q^2}\over F_2^N(x,Q^2)_{exp}}2x[1+R(x,Q^2)_{exp}]~,
\label{g1F2Rht}
\end{eqnarray}
where $g_1^N(x,Q^2)_{\rm LT}$ (N=p, n, d) is given by the leading twist
expression (\ref{g1partons}) in LO/NLO approximation including the
target mass corrections through the Nachtmann variable \cite{Nachtmann,
WMU}. In (\ref{g1F2Rht}) $h^N(x)$ are the dynamical $\tau=3$ and
$\tau=4$ HT corrections which  are extracted in a {\it
model independent way}. In our analysis their $Q^2$ dependence is
neglected. It is small and the accuracy of the present data does
not allow to determine it. For the unpolarized structure
functions $F_2^N(x,Q^2)_{exp}$ and $R(x,Q^2)_{exp}$ we have used
the NMC parametrization \cite{NMC} and the SLAC parametrization
$\rm R_{1998}$ \cite{R1998}, respectively. We will refer to these
as '($g_1+\rm HT$)' fits, and the set of polarized parton
densities extracted according to this approach - PD($g_1^{\rm
LT}+\rm HT$) or PD(Set 2).

As in our previous analyses \cite{LSS2001,LSSHT}, for the input
LO and NLO polarized parton densities at $Q^2_0=1~GeV^2$ we have
adopted a simple parametrization
\begin{eqnarray}
\nonumber
x\Delta u_v(x,Q^2_0)&=&\eta_u A_ux^{a_u}xu_v(x,Q^2_0),\\
\nonumber
x\Delta d_v(x,Q^2_0)&=&\eta_d A_dx^{a_d}xd_v(x,Q^2_0),\\
\nonumber
x\Delta s(x,Q^2_0)&=&\eta_s A_sx^{a_s}xs(x,Q^2_0),\\
x\Delta G(x,Q^2_0)&=&\eta_g A_gx^{a_g}xG(x,Q^2_0),
\label{inputPPD}
\end{eqnarray}
where on RHS of (\ref{inputPPD}) we have used the MRST98 (central
gluon) \cite{MRST98} and MRST99 (central gluon) \cite{MRST99}
parametrizations for the LO and NLO($\rm \overline{MS}$)
unpolarized densities, respectively. The normalization factors
$A_i$ in (\ref{inputPPD}) are fixed such that $\eta_{i}$ are the
first moments of the polarized densities. To fit better the data
in LO QCD, an additional factor $(1+ \gamma_v x)$ on the RHS is
used for the valence quarks. Bearing in mind that the light quark
sea densities $\Delta\bar{u}$ and $\Delta\bar{d}$ cannot, in
principle, be determined from the present inclusive data (in the
absence of polararized charged current neutrino experiments) we
have adopted the convention of a flavor symmetric sea
\begin{equation}
\Delta u_{sea}=\Delta\bar{u}=\Delta d_{sea}=\Delta\bar{d}= \Delta
s=\Delta\bar{s}. \label{SU3sea}
\end{equation}

The first moments of the valence quark densities $\eta_u$ and
$\eta_d$ are constrained by the sum rules
\begin{equation}
a_3=g_{A}=\rm {F+D}=1.2670~\pm~0.0035~~\cite{PDG}, \label{ga}
\end{equation}
\begin{equation}
a_8=3\rm {F-D}=0.585~\pm~0.025~~\cite{AAC00}, \label{3FD}
\end{equation}
where $a_3$ and $a_8$ are non-singlet combinations of the first
moments of the polarized parton densities corresponding to
$3^{\rm rd}$ and $8^{\rm th}$ components of the axial vector
Cabibbo current
\begin{equation}
a_3 = (\Delta u+\Delta\bar{u})(Q^2) - (\Delta
d+\Delta\bar{d})(Q^2)~, \label{a3ga}
\end{equation}
\begin{equation}
a_8 =  (\Delta u +\Delta\bar{u})(Q^2) + (\Delta d +
\Delta\bar{d})(Q^2) - 2(\Delta s+\Delta\bar{s})(Q^2)~. \label{a8}
\end{equation}

The polarized parton densities (\ref{inputPPD}) and
(\ref{SU3sea}) have to satisfy the positivity condition, which in
LO QCD implies:
\begin{equation}
\vert {\Delta f_i(x,Q^2_0)}\vert \leq f_i(x,Q^2_0),~~~~ \vert
{\Delta\bar{f_i}(x,Q^2_0)}\vert \leq \bar{f}_i(x,Q^2_0).
\label{pos}
\end{equation}

The constraints (\ref{pos}) are the consequence of a probabilistic
interpretation of the parton densities in the naive parton model,
which is still valid in LO QCD. Beyond LO the parton densities are
not physical quantities and the positivity constraints on the
polarized parton densities are more complicated. They follow from
the positivity condition for the polarized lepton-hadron
cross-sections $\Delta \sigma_i$ in terms of the unpolarized ones
($\vert {\Delta \sigma_i}\vert \leq \sigma_i$) and include also
the Wilson coefficient functions. It was shown \cite{AFR},
however, that for all practical purposes it is enough, at the
present stage, to consider LO positivity bounds for LO as well as
for for NLO parton densities, since NLO corrections are only
relevant at the level of accuracy of a few percent.

While in our previous NLO QCD analyses we have mainly used for the
unpolarized parton densities on the RHS of (\ref{pos}) the Barone
et al. parametrization \cite{Barone} we are here using the
MRST02(NLO) updated unpolarized parton densities \cite{MRST02} in
both $\rm \overline{MS}$ and JET schemes. The only significant
change in the MRST02 NLO partons, compared to
\vskip 0.6 cm
\begin{center}
\begin{tabular}{cl}
&{\bf Table 1.} The parameters of the {\bf Set 1} of NLO input
parton polarized densities \\ &PD($g_1^{\rm NLO}/F_1^{\rm NLO}$)
at $Q^2=1~GeV^2$ as obtained from the fits to the world \cite{EMC,world}
and \\&JLab \cite{JLab} data in the $\rm \overline{MS}$ and JET schemes.
The errors shown are total (statistical and \\& systematic).
The parameters marked by (*) are fixed by the sum rules
(\ref{ga}) and (\ref{3FD}).
\end{tabular}
\vskip 0.6 cm
\begin{tabular}{|c|c|c|c|c|c|c|} \hline
    Fit &~~~~$g_1^{\rm NLO}/F_1^{\rm NLO}(\rm \overline{MS})$~~~~
    &~~~~$g_1^{\rm NLO}/F_1^{\rm NLO}(\rm JET)$~~~~\\ \hline
 $\rm DF$        &  188~-~6            &     188~-~6 \\
 $\chi^2$        &  159.0              &     158.5    \\
 $\chi^2/\rm DF$ &  0.874              &     0.871   \\  \hline
 $\eta_u$        &~~0.926$^*$          &    $0.926^*$    \\
 $a_u$           &~~0.209~$\pm$~0.018~~&~~0.211~$\pm$~0.018  \\
 $\eta_d$        &-~0.341$^*$          &    $-0.341^*$      \\
 $a_d$           &~~0.072~$\pm$~0.068~~&~~0.083~$\pm$~0.058  \\
 $\eta_s$        &-~0.066~$\pm$~0.009~~&-~0.049~$\pm$~0.013  \\
 $a_s$           &~~0.649~$\pm$~0.117~~&~~0.779~$\pm$~0.214  \\
 $\eta_g$        &~~0.195~$\pm$~0.257~~&~~0.289~$\pm$~0.316  \\
 $a_g$           &~~2.575~$\pm$~1.729~~&~~0.000~$\pm$~0.686   \\ \hline
\end{tabular}
\end{center}
\vskip 0.6cm
those of MRST99 (which we have also used in some of
our analyses), is in the gluon, and especially in an increase in
the gluon density at high x. In the MRST02 fit to the world data
the authors no longer include prompt photon data due to
theoretical problems and possible inconsistencies between data
sets, and instead allow the high $x$ gluon to be determined by
the improved Tevatron jet data \cite{jetdata}, which considerably
improves the determination of the gluon. The use of new
positivity constraints leads to a significant change in the
polarized strange quark and gluon densities, the importance of
which will be discussed further.

\section{Results}

In this section we present the numerical results of our fits to
the world data \cite{EMC, world} on $g_1/F_1$ and $A_1$ including
JLab Hall A neutron data \cite{JLab}. The data used (188
experimental points) cover the following kinematic region:
\begin{equation}
0.005 \leq x \leq 0.75,~~~~~~1< Q^2 \leq 58~GeV^2~. \label{kinreg}
\end{equation}

\newpage
\vskip 0.6 cm
\begin{center}
\begin{tabular}{cl}
&{\bf Table 2.} The parameters of the {\bf Set 2} of LO, NLO($\rm
\overline{MS}$) and NLO(JET) input \\ &parton densities
PD($g_1^{\rm LT}+\rm HT$) at $Q^2=1~GeV^2$ as obtained from the
best $(g_1+\rm HT)$ \\ &fits to the world \cite{EMC, world} and JLab
\cite{JLab} data. The errors shown
are total (statistical and\\ &systematic). The parameters marked
by (*) are fixed. Note that the TMC are included\\
&in $(g_1)_{\rm LT}$.
\end{tabular}
\vskip 0.6 cm
\begin{tabular}{|c|c|c|c|c|c|c|} \hline
    Fit &~$(g_1)_{\rm LO}+h(x)/Q^2$~&~~$(g_1)_{\rm NLO(\rm \overline{MS})}+
h(x)/Q^2$~~
    &~~$(g_1)_{\rm NLO(\rm JET)}+h(x)/Q^2$ \\ \hline
 $\rm DF$        &  188~-~16          &     188~-~16   &  188~-~16  \\
 $\chi^2$        &  152.7             &      150.0     &    150.8 \\
 $\chi^2/\rm DF$ &  0.888             &      0.872     &    0.877 \\  \hline
 $\eta_u$        &~~0.926$^*$         &    $0.926^*$   &     $0.926^*$  \\
 $a_u$           &~~0.000~$\pm$~0.011~&~~0.244~$\pm$~0.036~&~~0.240~$\pm$~0.037 \\
 $\gamma_u$      &~~1.607~$\pm$~0.292~&    $0^*$           &   $0^*$ \\
 $\eta_d$        &-~0.341$^*$         &    $-0.341^*$      &   $-0.341^*$  \\
 $a_d$           &~~0.000~$\pm$~0.067~&~~0.123~$\pm$~0.134~&~~0.120~$\pm$~0.148  \\
 $\gamma_d$      &~~3.184~$\pm$~1.630~&    $0^*$           &   $0^*$  \\
 $\eta_s$        &-~0.075~$\pm$~0.010~&-~0.078~$\pm$~0.012~&-~0.064~$\pm$~0.017 \\
 $a_s$           &~~0.514~$\pm$~0.068~&~~0.629~$\pm$~0.090~&~~0.612~$\pm$~0.121 \\
 $\eta_g$        &~~$0.602^*$~        &~~0.348~$\pm$~0.345~&~~0.268~$\pm$~0.422 \\
 $a_g$           &~~$0.328^*$~        &~~1.980~$\pm$~1.359 &~~2.851~$\pm$~1.444 \\ \hline
 $x_i$           & \multicolumn{3}{|c|}{$h^p(x_i)~[GeV^2]$}   \\  \hline
  0.028          &~-0.003~$\pm$~0.036    &~~0.010~$\pm$~0.042 &~~0.024~$\pm$~0.039  \\
  0.100          &-~0.097~$\pm$~0.033    &-~0.043~$\pm$~0.034 &-~0.040~$\pm$~0.039 \\
  0.200          &-~0.164~$\pm$~0.032    &-~0.106~$\pm$~0.036 &-~0.108~$\pm$~0.038 \\
  0.350          &-~0.036~$\pm$~0.036    &-~0.016~$\pm$~0.038 &-~0.015~$\pm$~0.040 \\
  0.600          &~~0.032~$\pm$~0.019    &~~0.046~$\pm$~0.019 &~~0.048~$\pm$~0.019 \\ \hline
 $x_i$           &    \multicolumn{3}{|c|}{$h^n(x_i)~[GeV^2]$}  \\ \hline
  0.028          &~~0.204~$\pm$~0.078    &~~0.145~$\pm$~0.081 &~~0.161~$\pm$~0.081 \\
  0.100          &~~0.168~$\pm$~0.050    &~~0.192~$\pm$~0.047 &~~0.197~$\pm$~0.046 \\
  0.200          &~~0.023~$\pm$~0.058    &~~0.035~$\pm$~0.067 &~~0.032~$\pm$~0.070 \\
  0.325          &~~0.031~$\pm$~0.027    &~~0.019~$\pm$~0.031 &~~0.018~$\pm$~0.035 \\
  0.500          &~~0.031~$\pm$~0.015    &~~0.010~$\pm$~0.014 &~~0.011~$\pm$~0.016 \\ \hline
\end{tabular}
\end{center}
\vskip 2.0 cm

\newpage
The total (statistical and systematic) errors are taken into
account. The systematic errors are added quadratically.

We have determined from the data two sets of polarized parton
densities in both the $\rm \overline{MS}$ and the JET
factorization schemes: PD($g_1^{\rm NLO}/F_1^{\rm NLO}$) (see
Table 1) and PD($g_1^{\rm LT}+\rm HT$) (see Table 2). For the
second set of polarized parton densities a LO version
PD($g_1^{\rm LO}+\rm HT$) is also presented. The latter have been
extracted from the data using for the unpolarized parton densities
on the RHS of the positivity bounds (\ref{pos}) the LO MRST'01
ones (the MRST'02 LO set is essentially identical to the MRST'01
one and is not presented by the authors). Note once more that in
the LO '$g_1/F_1$' fit to the data the polarized parton densities
PD($g_1^{\rm LO}/F_1^{\rm LO}$) cannot be correctly determined
because the Callan-Gross relation $2xF_1(x,Q^2)_{\rm LO}=F_2(x,
Q^2)_{\rm LO}$ (used in the calculation of $F_1$) is strongly
broken (by up to 30 \%) for $x < 0.25$ and small $Q^2$. In other
words, to extract correctly the LO polarized parton densities
from the data, the second method of analysis have to be used,
i.e. the high twist corrections to $g_1$ have to be taken into
account. This observation is especially important for the
analysis of the semi-inclusive DIS data, where the LO QCD
approximation is mainly used.

\begin{figure}[ht]
\centerline{ \epsfxsize=2.2in\epsfbox{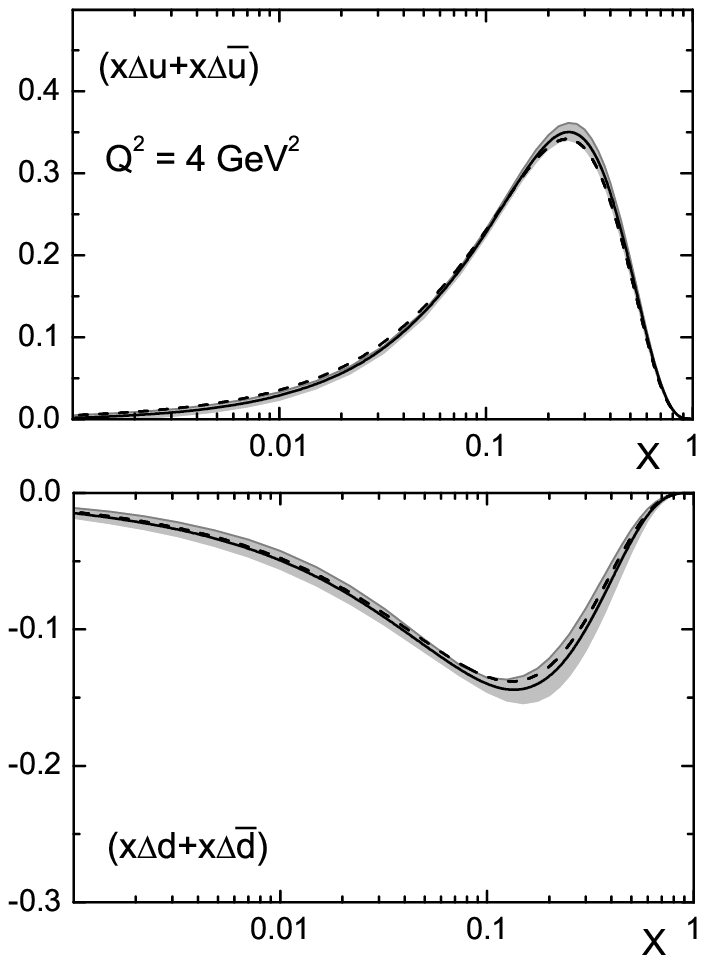}
\epsfxsize=2.2in\epsfbox{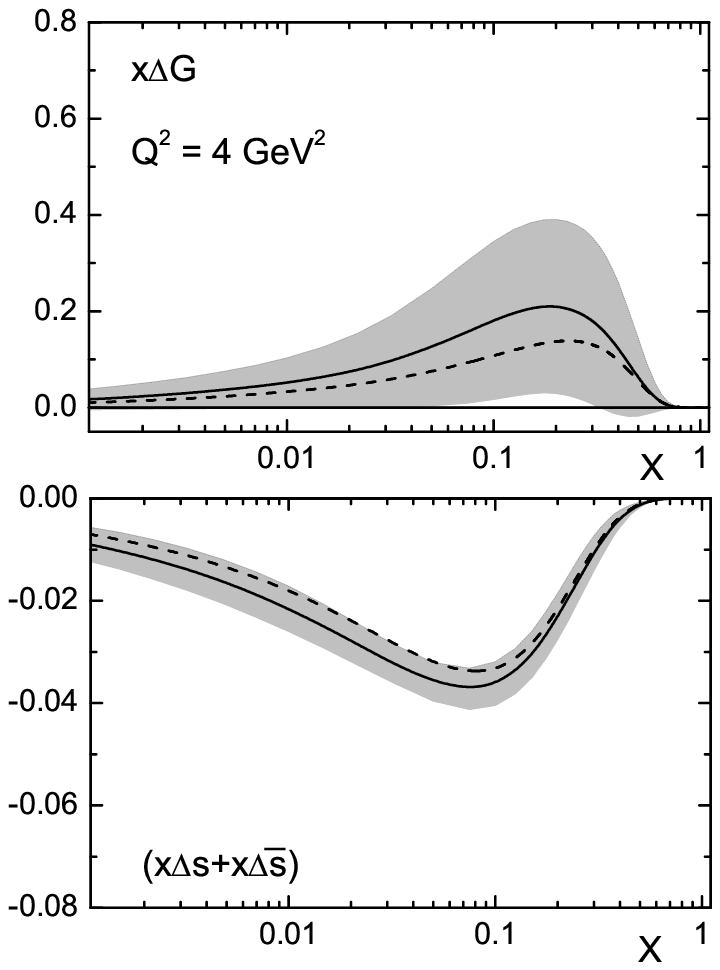} }
 \caption{NLO($\rm \overline{MS}$) polarized parton densities
PD($g_1^{\rm LT}+\rm HT$) (solid curves) together with their error
bands compared to PD($g_1^{\rm NLO}/F_1^{\rm NLO}$) (dashed
curves) at $Q^2=4~GeV^2$. \label{inter1}}
\end{figure}
In Fig. 1 we compare the NLO($\rm \overline{MS}$) polarized parton
densities PD($g_1^{\rm LT}+\rm HT$) with PD($g_1^{\rm
NLO}/F_1^{\rm NLO}$). As seen from Fig. 1 the two sets of
polarized parton densities are very close to each other,
especially for $u$ and $d$ quarks. This is a good illustration of
the fact that a fit to the $g_1$ data taking into account the
higher twist corrections to $g_1$ ($\chi^2_{\rm DF,NLO}=0.872$) is
equivalent to a fit of the data on $A_1(\sim g_1/F_1~)$ and
$g_1/F_1$ using for the $g_1$ and $F_1$ structure functions their
NLO leading twist expressions ($\chi^2_{\rm DF,NLO}=0.874$). In
other words, this fact confirms once more that the higher twist
corrections to $g_1$ and $F_1$ approximately cancel in the ratio
$g_1/F_1$. Nevertheless, we consider that the Set 2 of the
polarized parton densities PD($g_1^{\rm LT}+\rm HT$) is preferable
because using them and simultaneously extracted higher twist
corrections to $g_1$, the spin structure function $g_1$ can be
correctly calculated in the preasymptotic $(Q^2,~W^2)$ region too.

\begin{figure}[ht]
\centerline{ \epsfxsize=2.2in\epsfbox{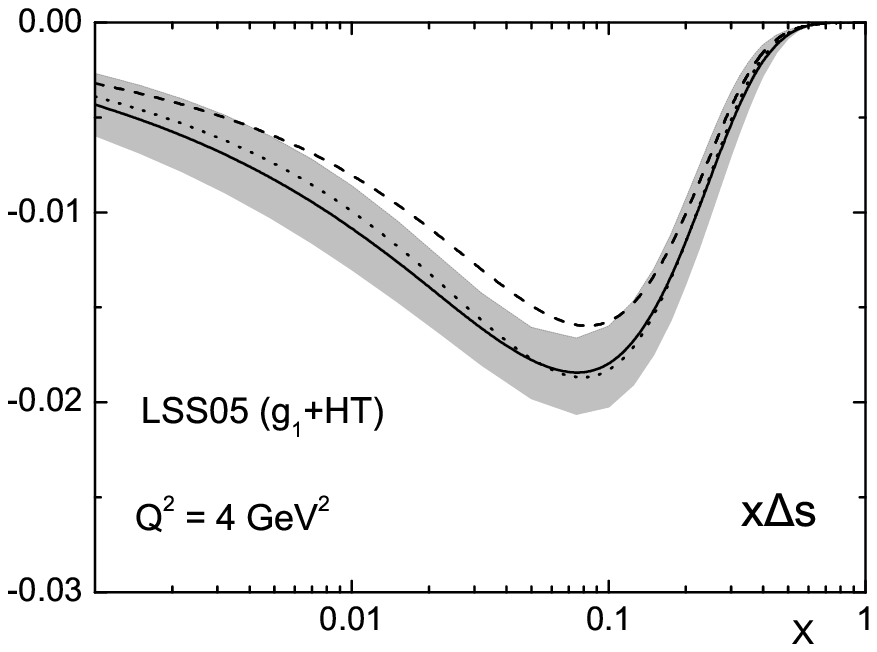}
\epsfxsize=2.2in\epsfbox{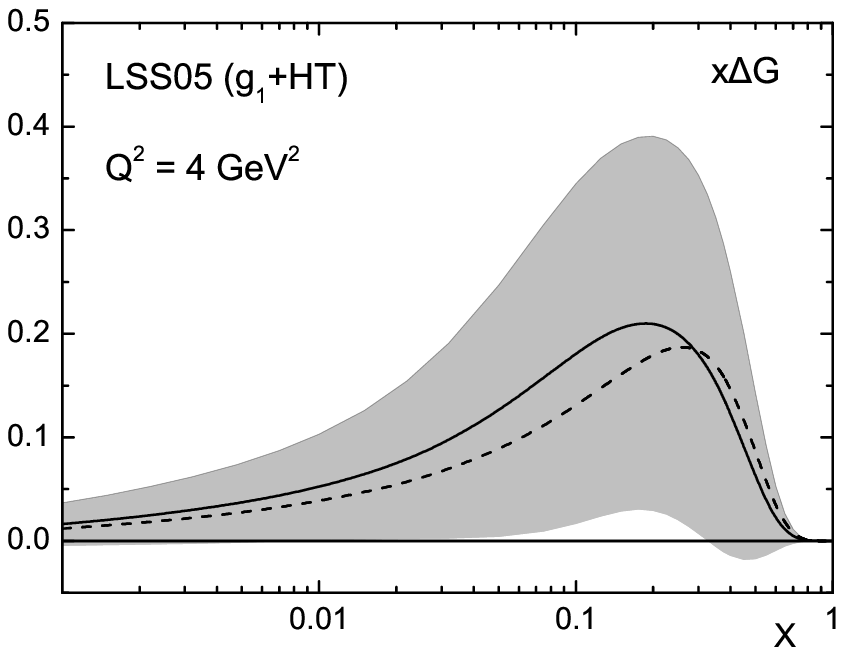} }
 \caption{
Comparison between NLO Set 2 polarized strange sea and gluon
densities at $Q^2=4~GeV^2$ in the $\rm \overline{MS}$ (solid
curves) and the JET scheme (dashed curves). Dot curve corresponds
to $\Delta s(x, Q^2)_{\rm JET \Rightarrow \overline {MS}}~$
obtained by the transformation rule (\ref{JETtoMS}) (see the
text). \label{inter2}}
\end{figure}

Let us briefly discuss the scheme dependence of our results.
Recall that according to perturbative QCD the NLO polarized
valence quarks and gluons should be the same [up to order of
${\cal O}(\alpha_{s}^3)$] in both the $\rm \overline{MS}$ and the
JET factorization schemes, while for the moments of the strange
sea quarks the following transformation rule is valid:
\begin{equation}
\nonumber \Delta s(n, Q^2)_{\rm \overline{MS}}=\Delta
s(n,Q^2)_{\rm JET}- {\alpha_s(Q^2)\over 2\pi n(n+1)}\Delta G(n,
Q^2)_{\rm JET}~. \label{JETtoMS}
\end{equation}

It is seen from the Table 1 and Table 2 that the values of
$~\chi^2/{\rm DF}(\rm \overline{MS})~$ and $~\chi^2/{\rm DF}(\rm
JET)$ coincide almost exactly for the $g_1/F_1$ as well as for the
$(g_1^{\rm LT}+\rm HT)$ fits, which is a good indication of the
stability of the analysis regardless of the scheme used. To
illustrate the factorization scheme dependence, the extracted
polarized PD($g_1^{\rm NLO}+\rm HT$) in the schemes under
consideration are compared in Fig. 2. Note that the valence
densities $\Delta u_v$ and $\Delta d_v$ in the $\rm
\overline{MS}$ and JET schemes are almost identical (see the
values of the parameters for the corresponding input parton
densities in Table 2) and in excellent agreement with what
follows from QCD.  So the corresponding curves are not shown in
Fig. 2. The extracted polarized gluons in the two schemes are
also well consistent within the errors (see Fig. 2). In Fig. 2 we
also show the polarized strange sea densities in both schemes,
determined directly from the fits and evolution equations, as
well as the strange sea $\Delta s(x, Q^2)_{\rm JET \Rightarrow
\overline {MS}}~$ obtained by the transformation rule
(\ref{JETtoMS}). It is seen that: i) at large $x$ $~\Delta s(x,
Q^2)_{\rm \overline {MS}}~$ is very close to $\Delta s(x,
Q^2)_{\rm JET}$ which is consistent with (\ref{JETtoMS}), since
large $n$ in the Mellin space corresponds to large Bjorken $x$,
and ii) $\Delta s(x, Q^2)_{\rm JET \Rightarrow \overline {MS}}~$
coincides very well with $\Delta s(x, Q^2)_{\rm \overline {MS}}$.
We have discussed here in detail the results on the set
PD($g_1^{\rm NLO}+\rm HT$). We have found the same conclusion for
the other set of polarized parton densities PD$(g_1^{\rm
NLO}/F_1^{\rm NLO})$. In conclusion, we have found that the
obtained numerical results are in a good agreement with the
perturbative QCD predictions.

Compared to our previous results \cite{LSS2001,LSSHT} we now
obtain smaller values for the gluon polarization (the first
moment of $\Delta G(x, Q^2))$, which leads to a smaller difference
between the values of the strange quark polarization (the first
moment of $\Delta s(x, Q^2)$) determined in the $\rm \overline
{MS}$ and the JET schemes, respectively. To illustrate this
tendency we present in Table 3 the corresponding values of the
first moments, $\Delta s$, $\Delta G$ and $\Delta \Sigma$, for
the LSS'01 and LSS'05(Set 1) sets of polarized parton densities.
Note that in the JET scheme the singlet  polarization $\Delta
\Sigma(Q^2)$ is a ${\it Q^2}$ {\it independent} quantity. Then, in
this scheme it is meaningful to directly interpret $\Delta
\Sigma$ as the contribution of the quark spins to the nucleon
spin and to compare its value obtained in the DIS region with the
predictions of the different (constituent, chiral, etc.) quark
models at low $Q^2(Q^2 \sim \Lambda ^2)$. Our new value of
$\Delta \Sigma_{\rm JET}= 0.29~\pm~0.08$ is smaller then the old
one and farther from the value 0.6 of $\Delta \Sigma$ at low
$Q^2$ region predicted in relativistic constituent quark models
\cite{RCQM}. Note, however, that if nonperturbative vacuum spin
effects are taken into account \cite{spinsea,instmodel}, the value
of $\Delta \Sigma$ at low $Q^2$ is expected to be smaller than 0.6.
Further theoretical and experimental investigation in both the
large and very low $Q^2$ regions, is needed to answer more
precisely the important question as to the fraction of the
nucleon spin carried by its quarks. \vskip 0.6cm
\begin{center}
\begin{tabular}{cl}
&{\bf Table 3.} First moments (polarizations) of LSS'01 and
LSS'05(Set 1) polarized \\ &parton densities at $Q^2 = 1~GeV^2$.
\end{tabular}
\vskip 0.4 cm
\begin{tabular}{|c|c|c|c|c|c|c|} \hline
 ~~Fit~~&${\Delta s(Q^2)}_{\rm \overline{MS}} $ & $\Delta s(Q^2)_{\rm JET}$ &
 $\Delta G(Q^2)_{\rm JET}$& $\Delta \Sigma_{\rm JET}$ \\ \hline
~LSS'01~&-0.065$~\pm$~0.016 &
 -0.035$~\pm~$0.010 & 0.68$~\pm~$0.32 & 0.37$~\pm~$0.07 \\ \hline
~LSS'05/Set 1~&-0.066$~\pm~$0.009 & -0.049$~\pm~$0.013 &
 0.29$~\pm~$0.32 & 0.29$~\pm~$0.08 \\ \hline
\end{tabular}
\end{center}
\vskip 0.4cm
\def\thefootnote{\dagger}
The extracted higher twist corrections to the proton and neutron
spin structure functions, $h^p(x)$ and $h^n(x)$, are shown in Fig.
3.\footnote{The moments of higher twist contribution to the proton
structure function $g_1$, including also in the analysis the data
on the resonance region, have been determined in the very recent
paper \cite{Osipenko}.} As seen from Fig. 3 the size of
\begin{figure}[th]
\centerline{ \epsfxsize=2.6in\epsfbox{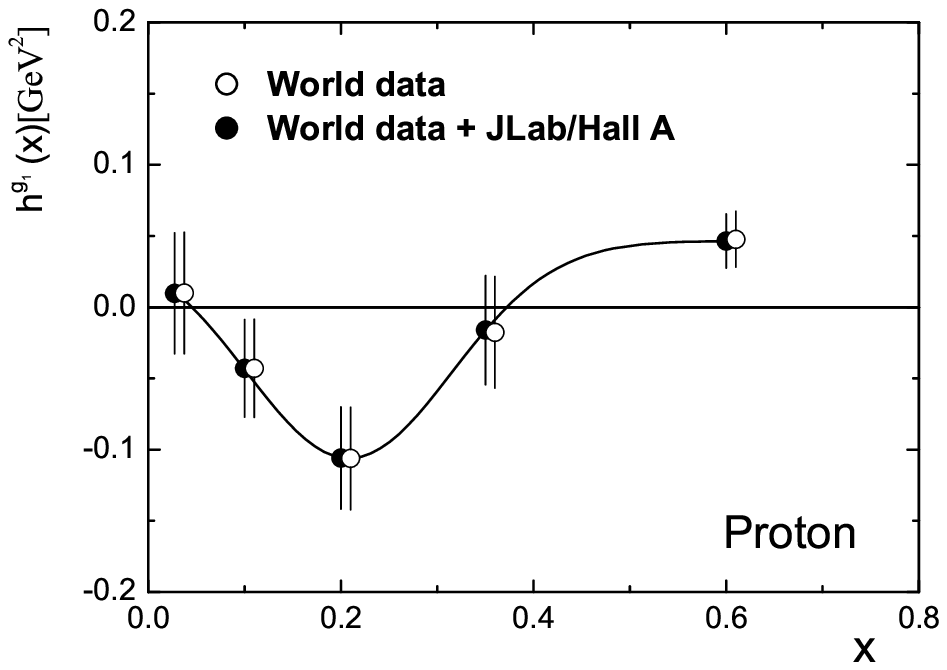}
\epsfxsize=2.6in\epsfbox{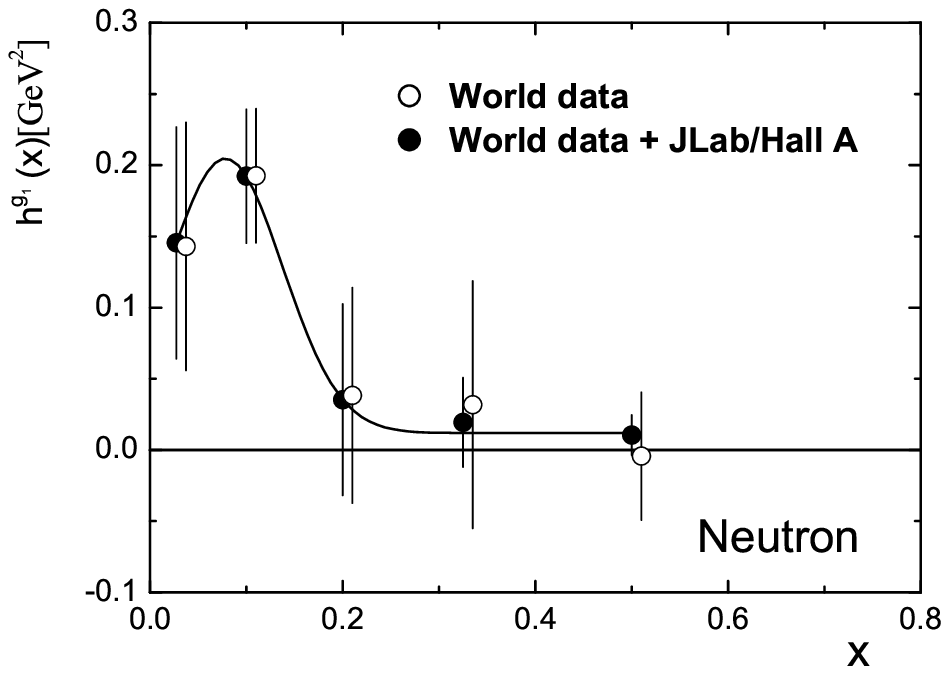} }
 \caption{
Higher twist corrections to the proton and neutron $g_1$ structure
functions extracted from the data on $g_1$ in NLO(${\rm \overline
{MS}}$) QCD approximation for $g_1(x,Q^2)_{\rm LT}$. The
parametrization (\ref{HTparam}) of the higher twist values is
also shown.\label{inter3}}
\end{figure}
the HT corrections is not negligible and their shape depends on
the target. In Fig. 3 our previous results on the higher twist
corrections to $g_1$ (before the JLab Hall A data were available)
are also presented. As seen from Fig. 3, thanks to the very
precise JLab Hall A data at large $x$ the higher twist corrections
to the neutron spin structure function are now much better
determined in this region. In Fig. 3 our parametrizations of the
values of higher twists for the proton and neutron targets

\begin{eqnarray}
\nonumber h^p(x)&=&0.0465 - {0.1913 \over \sqrt{\pi/2}}
exp[-2((x-0.2087)/0.2122)^2] \\
[2mm]
h^n(x)&=&0.0119 + {0.2420
\over \sqrt{\pi/2}} exp[-2((x-0.0783)/0.1186)^2] \label {HTparam}
\end{eqnarray}
are also shown. These should be helpful in a calculation of the
nucleon structure function $g_1$ for any $x$ and moderate $Q^2$ in
the experimental region, where the higher
\begin{figure}[ht]
\centerline{ \epsfxsize=2.3in\epsfbox{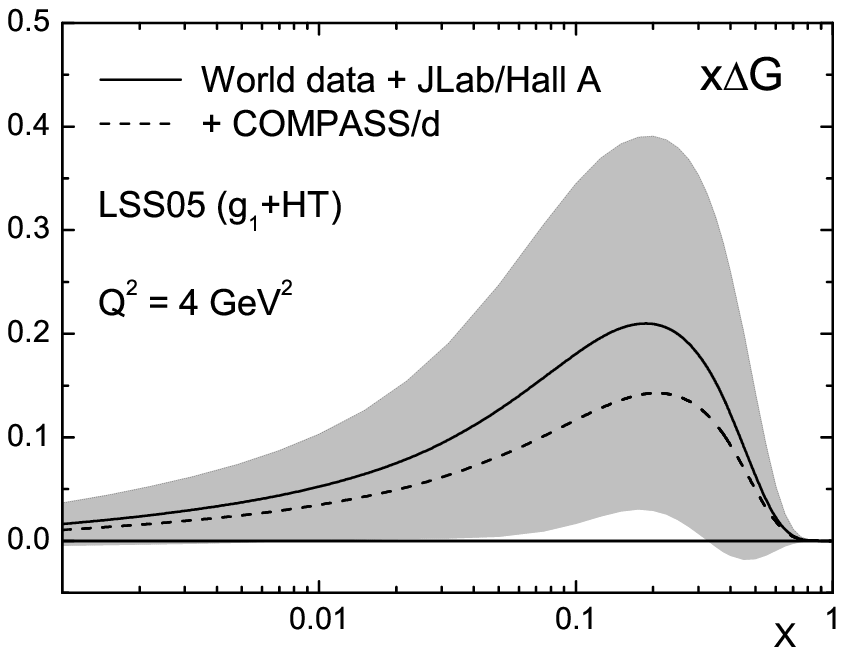}
\epsfxsize=2.3in\epsfbox{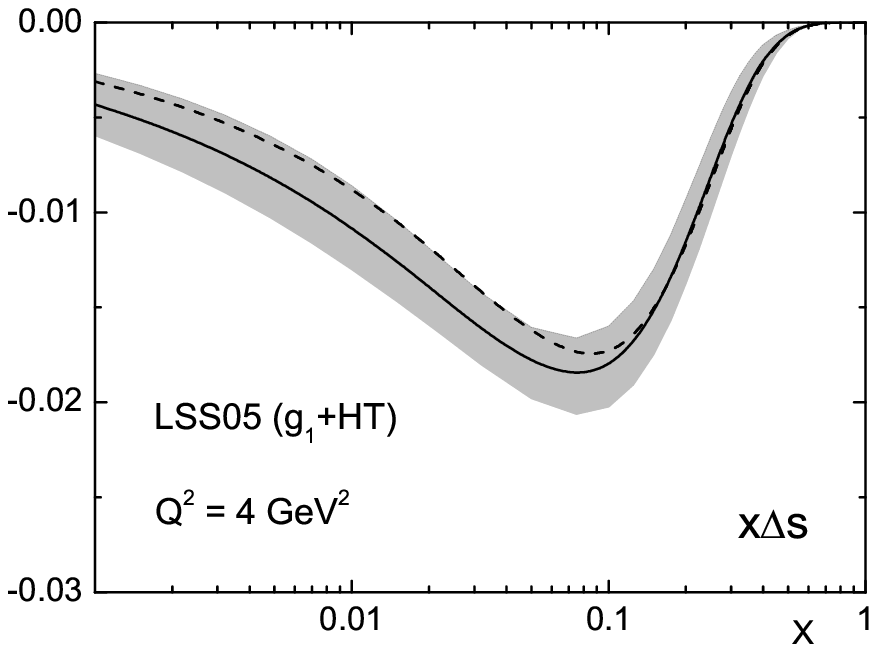} }
 \caption{
Effect of the COMPASS data on the Set 2 NLO($\rm \overline{MS}$)
polarized parton densities PD($g_1^{\rm NLO}+\rm HT$).
\label{inter4}}
\end{figure}
twist corrections are not negligible. The values of the higher
twist corrections to the proton and neutron $g_1$ structure
functions extracted in a model independent way from polarized DIS
data are in agreement with the QCD sum rule estimates
\cite{Balitsky:1990jb} as well as with the instanton model
predictions \cite{Balla:1997hf} but disagree with the renormalon
calculations \cite{HTrenorm}.

When this analysis was finished, the COMPASS Collaboration at CERN
reported new data on the longitudinal asymmetry $A_1^d$
\cite{COMPASS}. Their results improve considerably the
statistical accuracy on the small $x$ region $0.004 < x < 0.03$.
We have carried out a preliminary study ('$g_1+\rm HT$' fit in
the $\rm \overline{MS}$ scheme) to see whether the COMPASS data has
any significant effect on the result of our analysis.
The effect of the new
\begin{figure}[bht]
\centerline{ \epsfxsize=2.0in\epsfbox{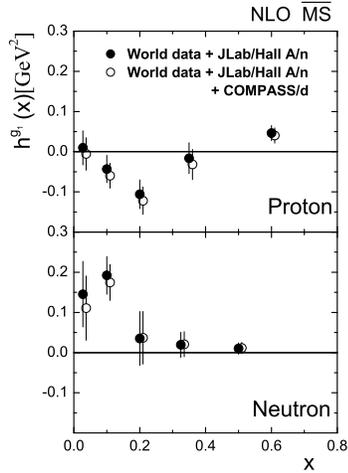}
 }
 \caption{
 Effect of the COMPASS data on the higher twist values.
 \label{inter5}}
\end{figure}
data on the polarized parton densities PD($g_1^{\rm NLO}+\rm HT$)
and the higher twist corrections is illustrated in Fig. 4 and Fig.
5. While the valence quarks densities $\Delta u_v$ and $\Delta
d_v$ do not change in the experimental region (for that reason
they are not shown in Fig. 4), the magnitudes of both the
polarized gluon and strange quark sea densities decrease, but the
corresponding curves lie within the error bands (see Fig. 4). The
impact of the new data on the values of higher twist corrections
is negligible (see Fig. 5). The new values are in good agreement
with the old ones although there is a tendency for the central
values for the proton target to be slightly lower than the old
ones. The central values of the HT correction for the neutron at
small $x$ are also slightly lower than the old ones.

\section{Impact of positivity constraints on polarized PD}

Let us consider now how the use of different positivity
constraints influences the results on the polarized parton
densities. In Fig. 6 we compare our new Set 1 of NLO($\rm
\overline {MS}$) polarized parton densities PD($g_1^{\rm
NLO}/F_1^{\rm NLO}$) with LSS'2001 parton densities \cite{LSS2001}
presented on the HEPDATA web site. Both sets are determined from
the data by the same method but using different positivity
constraints. (Note that the inclusion of the JLab data do not
influence the results on the polarized parton densities if the same
positivity constraints are used.) While
the new polarized PD($g_1^{\rm NLO}/F_1^{\rm
NLO}$) are compatible with the positivity bounds (\ref{pos})
imposed by the MRST'02 unpolarized parton densities \cite{MRST02},
those of the LSS'2001 set are limited by the Barone et al.
unpolarized parton densities \cite{Barone}. As seen from Fig. 6
the valence quark densities $\Delta u_v$ and $\Delta d_v$ of the
two sets are close to each other, while the polarized strange sea
quark and gluon densities are significantly different. This
comparison is a good illustration of the fact that the present
\begin{figure}[bht]
\centerline{ \epsfxsize=2.2in\epsfbox{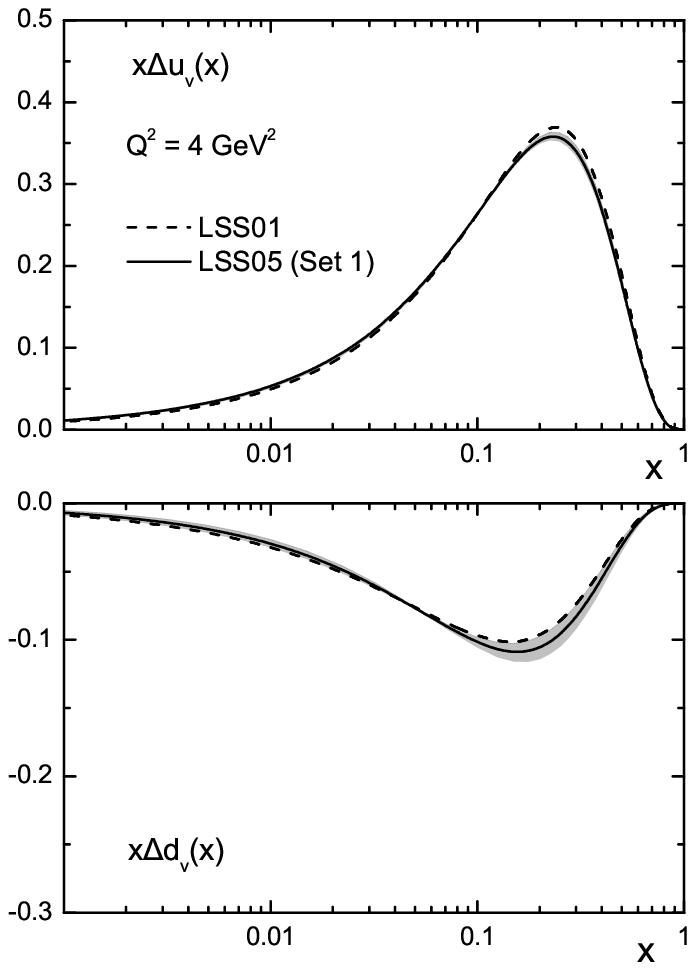}
\epsfxsize=2.2in\epsfbox{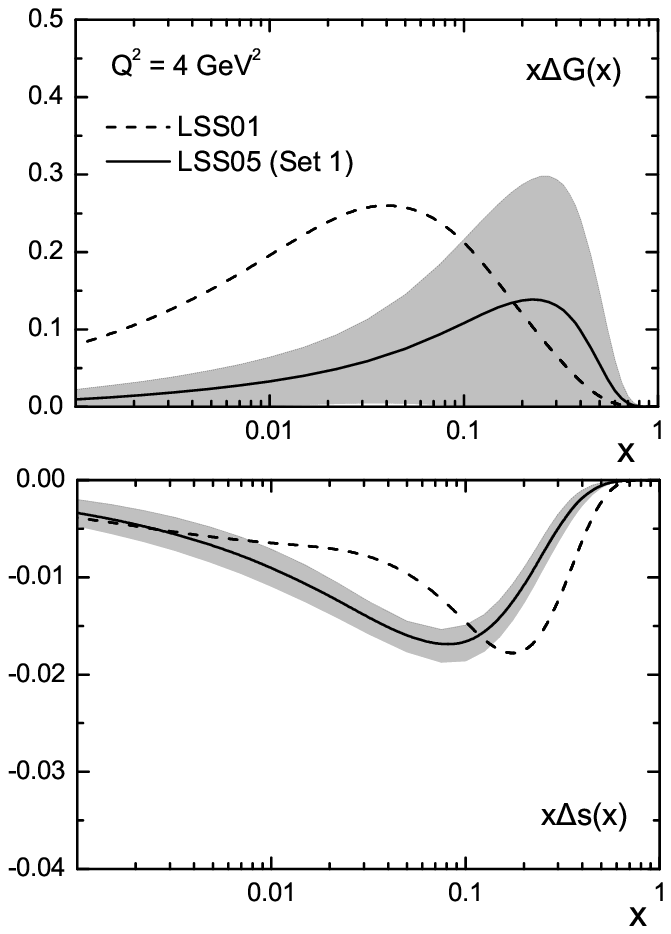} }
 \caption{
Comparison between our two sets of NLO(${\rm \overline {MS}}$)
polarized parton densities, LSS'01 and LSS'05(Set 1), at
$Q^2=4~GeV^2$.
 \label{inter6}}
\end{figure}
inclusive polarized DIS data allow a much better determination of
the valence quark densities (if SU(3) symmetry of the flavour
decomposition of the sea is assumed) than the polarized strange
quarks $\Delta s(x, Q^2)$ and the polarized gluons $\Delta G(x,
Q^2)$. This is especially true for the high $x$ region, where the
values of $\Delta s(x, Q^2)$ and $\Delta G(x, Q^2)$ are very small
and the precision of the data is not enough to extract them
correctly. That is why different unpolarized sea quark and gluon
densities (see Fig. 7) used on the RHS
\begin{figure}[t]
\centerline{ \epsfxsize=2.1in\epsfbox{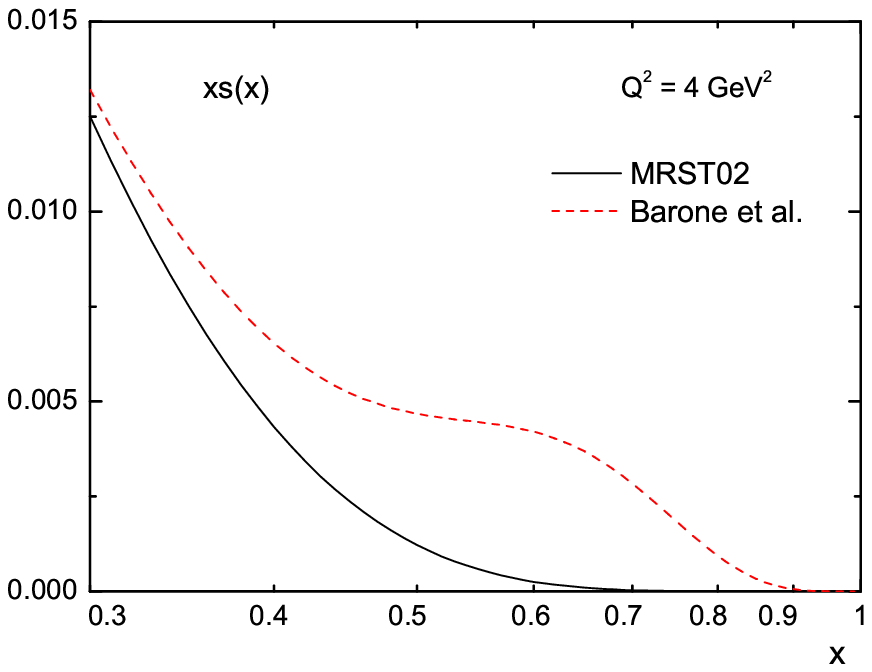}  \hskip 0.8 cm
\epsfxsize=2.0in\epsfbox{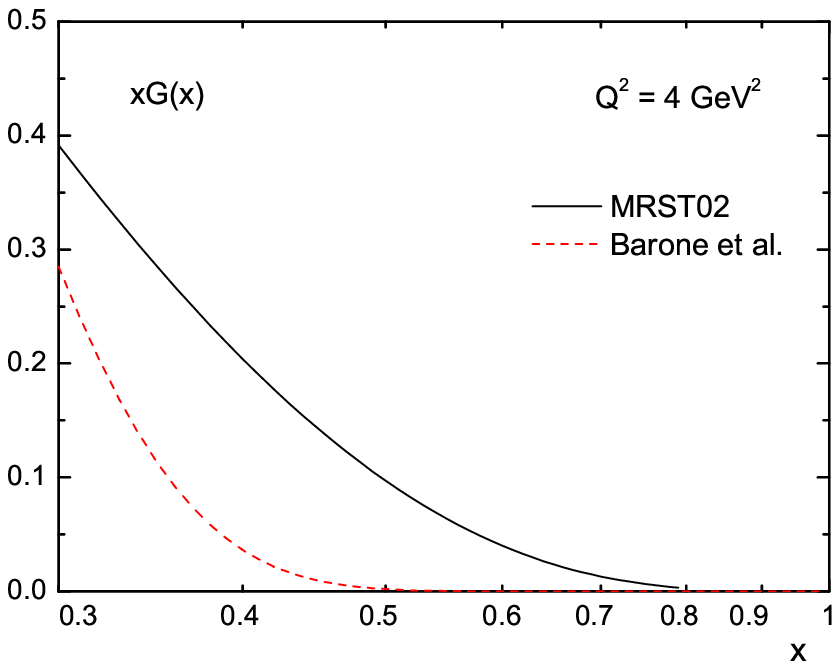} }
 \caption{Comparison between the NLO(${\rm \overline {MS}}$) unpolarized strange
quark sea and gluon densities determined by MRST'02 \cite{MRST02}
and Barone at al. \cite{Barone}.
 \label{inter7}}
\end{figure}
\begin{figure}[bh]
\centerline{ \epsfxsize=2.2in\epsfbox{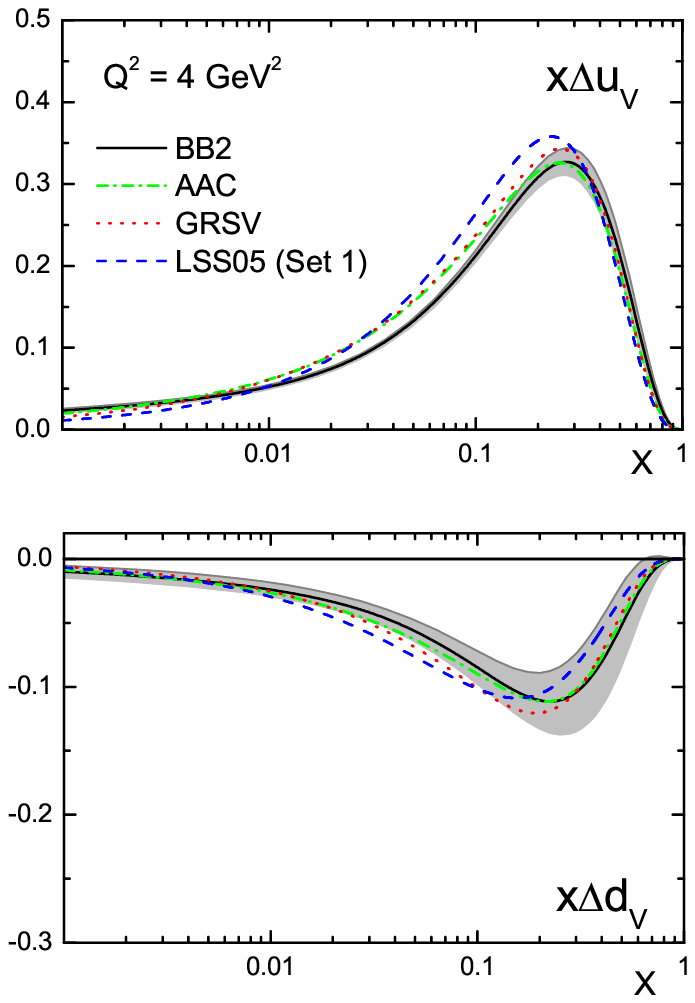}
 \epsfxsize=2.3in\epsfbox{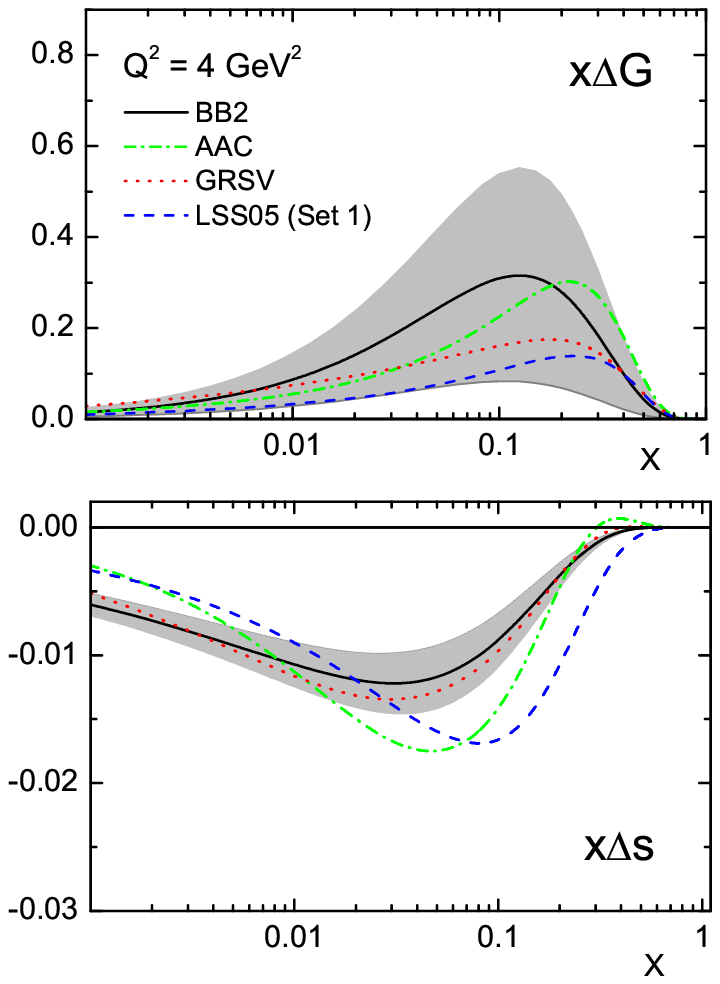}
 }
 \caption{
Comparison between our NLO(${\rm \overline {MS}}$) polarized
parton densities (Set 1) at $Q^2=4~GeV^2$ with those obtained by
GRSV ('standard scenario') \protect\cite{GRSV}, BB (ISET=4 or BB2)
\protect\cite{BB} and AAC (AAC03) \protect\cite{AAC04}.
 \label{inter8}}
\end{figure}
of the positivity constraints (\ref{pos}) are important and
crucial in determining $\Delta s(x, Q^2)$ and $\Delta G(x, Q^2)$
in this region. The more restrictive $s(x, Q^2)_{\rm MRST'02}$ at
high $x$ leads to a smaller value of $\vert {\Delta s(x,
Q^2)}\vert_{\rm LSS'05}$ in this region, while the smaller $G(x,
Q^2)_{\rm Bar.et.al}$ provides a stronger constraint on $\Delta
G(x, Q^2)_{\rm LSS'01}$ (see Fig. 6). To illustrate this fact once
more, we compare our new polarized parton densities PD($g_1^{\rm
NLO}/F_1^{\rm NLO}$) at $Q^2=4~GeV^2$
\begin{figure}[bht]
\centerline{ \epsfxsize=2.3in\epsfbox{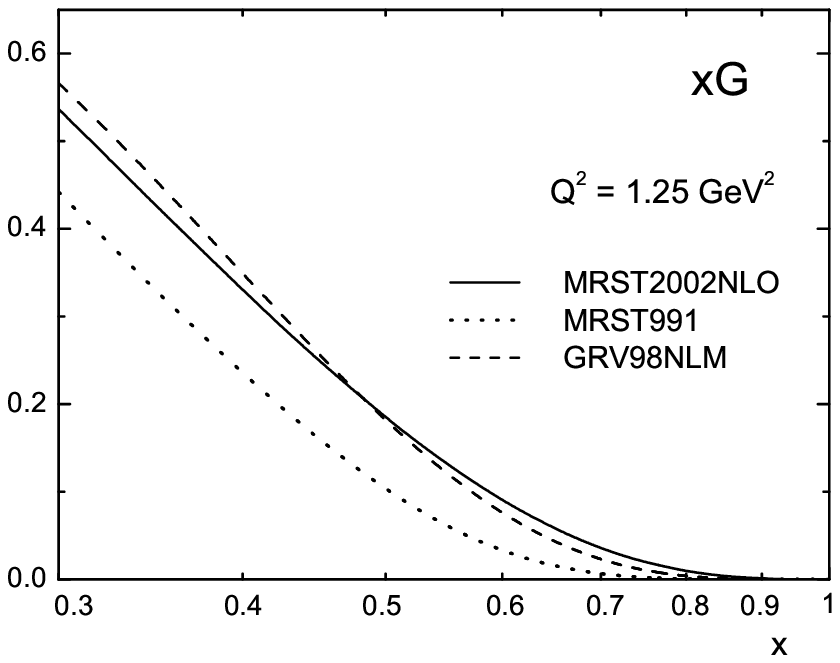}
\epsfxsize=2.3in\epsfbox{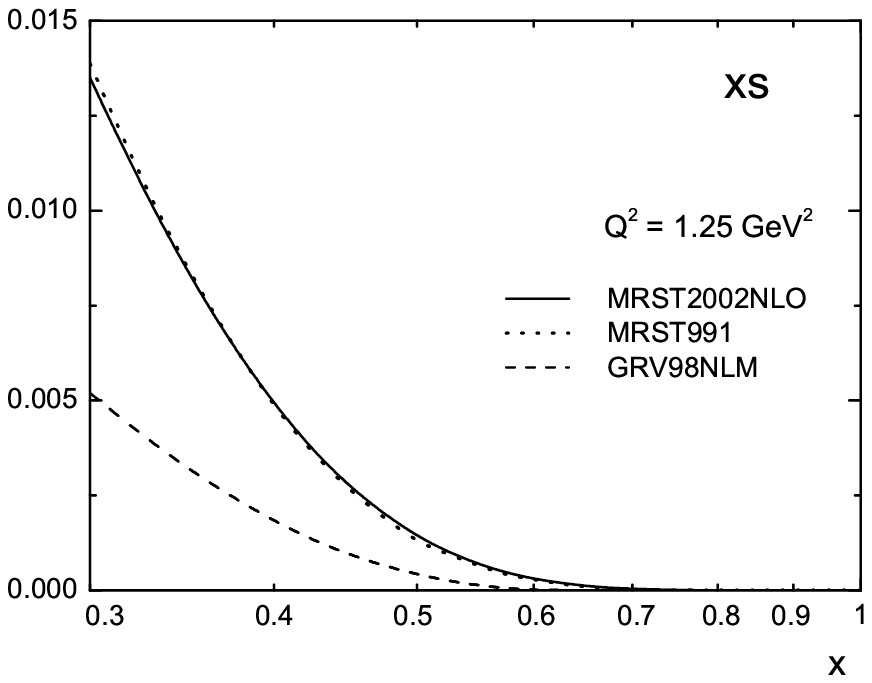} } \caption{ Comparison between
the NLO(${\rm \overline {MS}}$) unpolarized strange quark sea and
gluon densities determined by MRST \protect\cite{MRST99,MRST02}
and GRV \protect\cite{GRV}. \label{inter9} }
\end{figure}
with those obtained by GRSV \cite{GRSV}, Blumlein, Bottcher
\cite{BB} and the Asymmetry Analysis Collaboration (AAC)
\cite{AAC04} using almost the same set of data (see Fig. 8). Note
that all these groups have used the GRV unpolarized parton
densities \cite{GRV} for constraining their polarized parton
densities at large $x$. As seen from Fig. 9, in this $x$ region
the unpolarized GRV and MRST'02 gluons are practically the same,
while the magnitude of the unpolarized GRV strange sea quarks is
much smaller than that of MRST'02. Therefore, the GRV unpolarized
strange sea quarks provide a stronger constraint on the polarized
ones. The impact on the determination of the polarized strange sea
density is demonstrated in Fig. 10. (The GRV and MRST'02
unpolarized strange sea densities are also shown.) As a result the
magnitude of our polarized strange sea density $x\vert {\Delta
s(x, Q^2)}\vert$ is larger in the region $x> 0.1$ than those
obtained by the other groups. Note also that the magnitude of
$x\Delta s$ obtained by the GRSV and BB is smaller than that
determined by AAC. We consider the GRSV result to be a consequence
of the fact that in their analysis, the GRV positivity constraint
is imposed at lower value of $Q^2$: $Q^2=
{\mu}^2_{NLO}=0.4~GeV^2$, while AAC has used the same requirement
at $Q^2=1~GeV^2$. Finally, the different positivity conditions on
$\Delta s$ influence also the determination of the polarized gluon
density for larger $Q^2$ because the evolution in $Q^2$ mixes the
polarized sea quarks and gluons. To end this section we would like
to emphasize that for the adequate determination of polarized
strange quarks and gluons at large $x$, the role of the
corresponding unpolarized densities is very important. That is why
the latter have to be determined with good accuracy at large $x$
in the preasymptotic $(Q^2,~ W^2)$ region too. Usually the sets of
unpolarized parton densities, presented in the literature, are
extracted from the data on DIS using cuts in $Q^2$ and $W^2$
chosen in order to minimize the higher twist effects. In order to
use the densities for constraining the polarized parton densities
they have to be continued to the preasymptotic $(Q^2,~W^2)$
region. It is not obvious that the continued unpolarized parton
densities would coincide well with those obtained from the data in
the region $(Q^2>1~GeV^2,~ W^2 > 4~GeV^2)$ in the presence of the
HT corrections to unpolarized structure functions $F_1$ and $F_2$.
So, a QCD analysis of the unpolarized world data including the
preasymptotic $(Q^2,~W^2)$ region and taking into account HT
corrections is needed in order to extract correctly the
unpolarized parton densities in the preasymptotic
\begin{figure}[ht]
\centerline{ \epsfxsize=3.0in\epsfbox{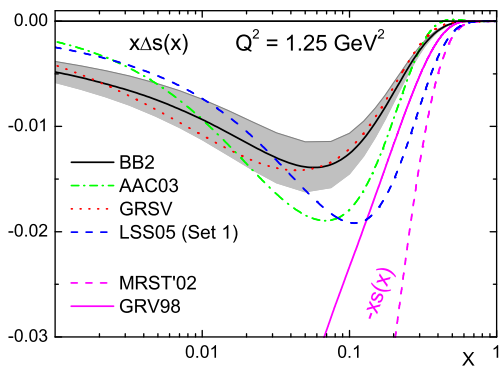}
 }
\caption{ Comparison between our NLO(${\rm \overline {MS}}$)
polarized strange sea quark density (Set 1) at $Q^2=1.25~GeV^2$
with those obtained by GRSV ('standard scenario')
\protect\cite{GRSV}, BB (ISET=4 or BB2) \protect\cite{BB} and AAC
(AAC03) \protect\cite{AAC04}. The unpolarized MRST02 and GRV98
strange sea quark densities are also shown.
\label{inter10}}
\end{figure}
region. Our arguments for the need for a precise determination of
the unpolarized densities of strange quarks and gluons in both the
asymptotic and preasymptotic regions in $Q^2$ and $W^2$, coming
from spin physics, could be considered as additional to those
discussed in the recent paper \cite{Forte}.

\section{Conclusions}

{\it i}) We have re-analyzed the world data on inclusive polarized
deep inelastic lepton-nucleon scattering in leading and
next-to-leading order of QCD, adding to the old set of data the
very precise JLab Hall A neutron data. Compared to our previous
analyses new positivity constraints on the polarized parton
densities have been used. The latter reflect mainly the better
determination of unpolarized gluon density, especially at high
$x$. Two new sets of NLO polarized parton densities in the JET and
$\rm \overline{MS}~$ factorization schemes as well as LO
polarized patron densities have been extracted from the data
using different methods of analysis. The NLO polarized parton
densities determined in the two schemes are in a good agreement
with the pQCD predictions.

{\it ii}) The impact of positivity constraints on the polarized
parton densities has been studied. Special attention has been
paid to the role of positivity constraints in determining the
polarized strange quark and gluon densities, which are not well
determined from the present inclusive DIS data. For that reason
the effect of the positivity conditions used to constrain them is
very important. On other hand, the different sets of unpolarized
parton densities needed to impose the positivity bounds
(\ref{pos}) are usually determined in the DIS region and their
continuation to the preasymptotic region via the evolution DGLAP
equations sometimes leads to very different behaviour of the
unpolarized parton densities belonging to the different sets. In
particular, it is demonstrated that the use of MRST'02 and GRV98
NLO($\rm \overline{MS}~$) unpolarized strange quark densities in
the RHS of the positivity condition (\ref{pos}) leads to a
significant difference of the extracted polarized strange sea
density, especially for $x > 0.05$. So, a more precise
determination of the unpolarized parton densities in the
preasymptotic region is very important for a better determination
of the polarized ones, especially of the polarized strange quark
sea and gluon densities, which are weakly constrained by the
present experimental data.

{\it iii}) It was demonstrated that in the fit to the $g_1$ data
the higher twist corrections to $g_1$ are important and have to be
taken into account. It was also shown that the values of higher
twist corrections to the neutron spin structure function at high
$x$ are determined much more precisely when the JLab Hall A data
are used in the analysis.

{\it ~iv}) Finally, the use of the very recent COMPASS data on
inclusive asymmetry $A_1^d$ has little or no effect on the
polarized valence densities, but the magnitudes of
the strange quark and gluon densities decrease.\\

\vskip 6mm { \bf Acknowledgments} \vskip 4mm This research was
supported by the JINR-Bulgaria Collaborative Grant, by the RFBR
(No 05-01-00992, 03-02-16816) and by the Bulgarian National
Science Foundation under Contract Ph-1010.

\vskip 25mm {\Large \bf Appendix} \vskip 6mm For practical
purposes we present here explicitly our Set 1 and Set 2 of
polarized parton densities at $Q^2=1~\rm GeV^2$. The polarized
valence quark
densities correspond to SU(3) flavour symmetric sea.  \\

LSS'05 ({\bf Set 1}) - NLO($\rm \overline {MS}$) PD($g_1/F_1$):
\setcounter{equation}{0}
\renewcommand\theequation{A.\arabic{equation}}
\begin{eqnarray}
\nonumber x\Delta
u_v(x)&=&~~0.4621~x^{0.6258}~(1-x)^{3.428}~(~1+2.179~x^{1/2}+
14.57~x~)~,\\
\nonumber x\Delta
d_v(x)&=&-0.02257~x^{0.3429}~(1-x)^{3.864}~(~1+35.47~x^{1/2}+
28.97~x~)~,\\
\nonumber x\Delta
s(x)&=&-0.02520~x^{0.3669}~(1-x)^{7.649}~(~1+3.656~x^{1/2}+
19.50~x~)~,\\
x\Delta G(x)&=&~~532.3~x^{3.544}~(1-x)^{6.879}~(~1-3.147~x^{1/2}+
3.148~x~)~. \label{Set1NLOMS}
\end{eqnarray}
\vskip 1cm

LSS'05 ({\bf Set 1}) - NLO(JET) PD($g_1/F_1$):
\begin{eqnarray}
\nonumber x\Delta
u_v(x)&=&~~0.4635~x^{0.6272}~(1-x)^{3.428}~(~1+2.179~x^{1/2}+
14.57~x~)~,\\
\nonumber x\Delta
d_v(x)&=&-0.02322~x^{0.3539}~(1-x)^{3.864}~(~1+35.47~x^{1/2}+
28.97~x~)~,\\
\nonumber x\Delta
s(x)&=&-0.02867~x^{0.4973}~(1-x)^{7.649}~(~1+3.656~x^{1/2}+
19.50~x~)~,\\
x\Delta
G(x)&=&~~5.080~x^{0.9692}~(1-x)^{6.879}~(~1-3.147~x^{1/2}+
3.148~x~)~.
\label{Set1NLOJET}
\end{eqnarray}
\vskip 1cm

LSS'05 ({\bf Set 2}) - LO PD($g_1 + \rm HT$):
\begin{eqnarray}
\nonumber x\Delta
u_v(x)&=&~~0.1761~x^{0.3012}~(1-x)^{3.177}~(1+1.607~x)(~1-0.4085~x^{1/2}+
17.60~x~)~,\\
\nonumber x\Delta
d_v(x)&=&-0.00807~x^{0.1535}~(1-x)^{3.398}~(1+3.184~x)(~1+37.25~x^{1/2}+
31.14~x~)~,\\
\nonumber x\Delta
s(x)&=&-0.04464~x^{0.3239}~(1-x)^{8.653}~(~1-0.9052~x^{1/2}+
11.53~x~)~,\\
x\Delta G(x)&=&~~1.164~x^{0.4536}~(1-x)^{5.511}~(~1-4.255~x^{1/2}+
7.274~x~)~. \label{Set2LO}
\end{eqnarray}
\vskip 1cm

LSS'05 ({\bf Set 2}) - NLO($\rm \overline {MS}$) PD($g_1 + \rm
HT$):
\begin{eqnarray}
\nonumber x\Delta
u_v(x)&=&~~0.4958~x^{0.6606}~(1-x)^{3.428}~(~1+2.179~x^{1/2}+
14.57~x~)~,\\
\nonumber x\Delta
d_v(x)&=&-0.02567~x^{0.3936}~(1-x)^{3.864}~(~1+35.47~x^{1/2}+
28.97~x~)~,\\
\nonumber x\Delta
s(x)&=&-0.02756~x^{0.3472}~(1-x)^{7.649}~(~1+3.656~x^{1/2}+
19.50~x~)~,\\
x\Delta G(x)&=&~~421.9~x^{2.949}~(1-x)^{6.879}~(~1-3.147~x^{1/2}+
3.148~x~)~. \label{Set2NLOMS}
\end{eqnarray}
\vskip 1cm

LSS'05 ({\bf Set 2}) - NLO(JET) PD($g_1 + \rm HT$):
\begin{eqnarray}
\nonumber x\Delta
u_v(x)&=&~~0.4924~x^{0.6571}~(1-x)^{3.428}~(~1+2.179~x^{1/2}+
14.57~x~)~,\\
\nonumber x\Delta
d_v(x)&=&-0.02549~x^{0.3909}~(1-x)^{3.864}~(~1+35.47~x^{1/2}+
28.97~x~)~,\\
\nonumber x\Delta
s(x)&=&-0.02115~x^{0.3301}~(1-x)^{7.649}~(~1+3.656~x^{1/2}+
19.50~x~)~,\\
x\Delta G(x)&=&~~1009.5~x^{3.820}~(1-x)^{6.879}~(~1-3.147~x^{1/2}+
3.148~x~)~. \label{Set2NLOJET}
\end{eqnarray}


\vskip 1cm

\begin{thebibliography}{99}

\bibitem{EMC}
EMC, J. Ashman {\it et al.}, Phys. Lett. {\bf B 206} (1988) 364;
Nucl. Phys. {\bf B 328} (1989) 1.

\bibitem{JET}
R. D. Carlitz, J. C. Collins and A.H. Mueller, Phys. Lett. {\bf B
214} (1988) 229; M. Anselmino, A. V. Efremov and E. Leader, Phys.
Rep. {\bf 261} (1995) 1; H.-Y. Cheng, Int. J. Mod. Phys. {\bf A
11} (1996) 5109; D. M\"{u}ller and O. V. Teryaev, Phys. Rev. {\bf
D 56} (1997) 2607.

\bibitem{world} SLAC E142 Coll., P.L. Anthony {\it et al.},
Phys. Rev. {\bf D 54} (1996) 6620; SLAC/E154 Coll., K. Abe {\it et
al.}, Phys. Rev. Lett. {\bf 79} (1997) 26; SMC, B. Adeva {\it et
al.}, Phys. Rev. {\bf D 58} (1998) 112001; HERMES, K. Ackerstaff
{\it et al.}, Phys. Lett. {\bf B 404} (1997) 383; {\it ibid} {\bf
B 442} (1998) 484; SLAC E143 Coll., K. Abe {\it et al.}, Phys.
Rev. {\bf D 58} (1998) 112003; SLAC/E155 Coll., P.L. Anthony {\it
et al.}, Phys. Lett. {\bf B 463} (1999) 339, {\it ibid} {\bf B
493} (2000) 19.

\bibitem{JLab}
JLab/Hall A Coll., X. Zheng {\it et al.}, Phys. Rev. Lett. {\bf
92} (2004) 012004.

\bibitem{LSS2001}
E. Leader, A.V. Sidorov and D.B. Stamenov, Eur. Phys. J. {\bf C
23} (2002) 479.

\bibitem{LSSHT}
E. Leader, A.V. Sidorov and D.B. Stamenov, Phys. Rev. {\bf D 67}
(2003) 074017.

\bibitem{TB}
J. Blumlein and A. Tkabladze, Nucl. Phys. Proc. Suppl. {\bf 79}
(1999) 541; S. Simula, M. Osipenko, G. Ricco and M. Taiuti, Phys.
Rev. {\bf D 65} (2002) 034017.

\bibitem{WW}
S. Wandzura and F. Wilczek, Phys. Lett. {\bf B 72} (1977) 195.

\bibitem{LomConf}
E. Leader, A.V. Sidorov and D.B. Stamenov, in {\it Particle
Physics at the Start of the New Millennium}, edited by A.I.
Studenikin, World Scientific, Singapore, May 2001, p. 76. ({\it
Proceedings of the 9th Lomonosov Conference on Elementary
Particle Physics, Moscow, Russia, 20-26 Sep 1999}).

\bibitem{newHTA1}
E. Leader, A. V. Sidorov and D. B. Stamenov, in {\it Deep
Inelastic Scattering DIS2003}, edited by V.Kim and L.Lipatov,
PNPI RAS, 2003, pp. 790-794, hep-ph/0309048.

\bibitem{Nachtmann}
O. Nachtmann, Nucl. Phys. {\bf B 63} (1973) 237.

\bibitem{WMU}
S. Wandzura, Nucl. Phys. {\bf B 122} (1977) 412; S. Matsuda and
T. Uematsu, Nucl. Phys. {\bf B 168} (1980) 181.

\bibitem{NMC}
NMC Coll., M. Arneodo {\it et al.}, Phys. Lett. {\bf B 364} (1995)
107.

\bibitem{R1998}
SLAC/E143 Coll., K. Abe {\it et al.}, Phys. Lett. {\bf B 452}
(1999) 194.

\bibitem{MRST98}
A.D. Martin, R.G. Roberts, W.J. Stirling and R.S. Thorne, Eur.
Phys. J. {\bf C 4} (1998) 463.

\bibitem{MRST99}
A.D. Martin, R.G. Roberts, W.J. Stirling and R.S. Thorne, Eur.
Phys. J. {\bf C 14} (2000) 133.

\bibitem{PDG}
Particle Data Group, Eur. Phys. J. {\bf C 15} (2000) 695.

\bibitem{AAC00}
Asymmetry Analysis Collaboration, Y. Goto {\it et al.}, Phys.
Rev. {\bf D 62} (2000) 034017.

\bibitem{AFR}
G. Altarelli, S. Forte and G. Ridolfi, Nucl. Phys. {\bf B 534}
(1998) 277; S. Forte, M. L. Mangano and G. Ridolfi, Nucl. Phys.
{\bf B 602} (2001) 585.

\bibitem{Barone}
V. Barone, C. Pascaud and F. Zomer, Eur. Phys. J. {\bf C 12}
(2000) 243.

\bibitem{MRST02}
A.D. Martin, R.G. Roberts, W.J. Stirling and R.S. Thorne, Eur.
Phys. J. {\bf C 28} (2003) 455.

\bibitem{jetdata}
D0 Coll., B. Abbot {\it et al.}, Pys. Rev. Lett. {\bf 86} (2001)
1707; CDF Coll., T. Affolder {\it et al.}, Phys. Rev. {\bf D 64}
(2001) 032001.

\bibitem{RCQM}
A.W. Shreiber and A.W. Thomas, Phys. Lett. {\bf B 215} (1988) 141;
R.D. Jaffe and A. Manohar, Nucl. Phys. {\bf B 337} (1990) 509.

\bibitem{spinsea}
S.N. Shore and G. Veneziano, Phys. Lett. {\bf B 244} (1990) 75; S. Forte
and E.V. Shuryak, Nucl. Phys. {\bf B 357} (1991) 153.

\bibitem{instmodel}
A.E. Dorokhov, Czech. J. Phys. {\bf 52} (2002) c79; A.E. Dorokhov,
N.I. Kochelev and Yu.A. Zubov, Int. Journ. Mod. Phys. {\bf A 8}
(1993) 603; A.E. Dorokhov, N.I. Kochelev, Phys. Lett. {\bf B 304}
(1993) 167.
%hep-ph/0112332

\bibitem{Osipenko}
M. Osipenko {\it et al.}, Phys. Rev. {\bf D 71} (2005) 054007.
%hep-ph/0503018.

\bibitem{Balitsky:1990jb}
I.I. Balitsky, V.M. Braun and A.V. Kolesnichenko, Phys. Lett.
{\bf B 242} (1990) 245. Erratum {\it ibid} {\bf B318} (1993) 648;
E. Stein {\it et al.}, Phys. Lett. {\bf B 353} (1995) 107.

\bibitem{Balla:1997hf}
J. Balla, M.V. Polyakov and C. Weiss, Nucl. Phys. {\bf B 510}
(1998) 327; A.V. Sidorov and C. Weiss, hep-ph/0410253.

\bibitem{HTrenorm}
E. Stein, Nucl. Phys. Proc. Suppl. {\bf 79} (1999) 567.

\bibitem{COMPASS}
COMPASS Coll., E.S. Ageev {\it et al.}, hep-ex/0501073.

\bibitem{GRSV}
M. Gl\"{u}ck, E. Reya, M. Stratmann and W. Vogelsang, Phys. Rev.
{\bf D 63} (2001) 094005.

\bibitem{BB}
J. Blumlein, H. Bottcher, Nucl. Phys. {\bf B 636} (2002) 225.

\bibitem{AAC04}
Asymmetry Analysis Collaboration, M. Hirai {\it et al.}, Phys.
Rev. {\bf D 69} (2004) 054021.

\bibitem{GRV}
M. Gl\"{u}ck, E. Reya and A. Vogt, Eur. Phys. J. {\bf C 5} (1998)
461.

\bibitem{Forte}
S. Forte, hep-ph/0502073.

\end{thebibliography}
\end{document}